# A New Class of Probability Distributions for Describing the Spatial Statistics of Area-averaged Rainfall

Prasun K. Kundu[1] and Ravi K. Siddani[2]

[1]*Joint Center for Earth Systems Technology (JCET), University of Maryland Baltimore County, Baltimore, MD, and NASA Goddard Space Flight Center, Greenbelt, MD*

[2]*Department of Mathematics and Statistics, University of Maryland Baltimore County, Baltimore, MD*

## ABSTRACT

Rainfall exhibits extreme variability at many space and time scales and calls for a statistical description. Based on an analysis of radar measurements of precipitation over the tropical oceans, we introduce a new probability law for the area-averaged rain rate constructed from the class of log-infinitely divisible distributions that accurately describes the frequency of the most intense rain events. The dependence of its parameters on the spatial averaging length $L$ allows one to relate spatial statistics at different scales. In particular, it enables us to explain the observed power law scaling of the moments of the data and successfully predicts the continuous spectrum of scaling exponents expressing multiscaling characteristics of the rain intensity field.

## 1. Introduction.

The intrinsic unpredictability of rainfall intensity, which varies in an irregular manner in time and space, makes it natural to seek a statistical description in terms of an underlying probability distribution. Since in the final analysis rainfall merely consists of a collection of falling raindrops of various sizes, the instantaneous point rain rate field is a highly singular mathematical quantity that becomes accessible to large-scale observation only through space and/or time averaging. In this respect, radar remote-sensing measurements and the traditional rain gauge measurements probe the rain rate field in two distinct regimes. Radar observations are a convenient way to measure the near-instantaneous rain rate averaged over a certain area (typically a few kilometers in size) determined by the intrinsic spatial resolution of the experimental setup. A sequence of gridded radar-derived rain maps over a larger area, typically a few hundred kilometers in size, can be utilized to study statistical properties of area-averaged rain. On the other hand, rain gauges measure time-averaged rain rate over a very small area (of the order of a few tens of centimeters) that can be well approximated as a point. In this paper we seek a theoretical description of the statistical properties of the spatial structure of rain and introduce a new family of probability distributions for describing rainfall statistics, focusing on the spatial statistics of area-averaged rain rate derived from radar remote sensing.

An interesting aspect of rainfall statistics is that they depend in a non-trivial manner on the length and time scales over which rain rate is averaged. There have been a number of different theoretical approaches to modeling the scale dependence of rain statistics. Inspired by the statistical theory of fully developed turbulence, phenomenological models based on multiplicative random cascade process have been proposed for the spatial statistics [*Lovejoy and Schertzer* 1985, *Schertzer and Lovejoy* 1987; *Gupta and Waymire* 1990, 1993], the temporal statistics [*Veneziano et al.* 1996, *Menabde et al.* 1997, *Olsson and Burlando* 2002] and the full space-time statistics [*Marsan et al.* 1996; *Over and Gupta* 1996; *Seed et al.* 1999]. These models aim to capture the power law dependence of the moments of the area- or time- averaged rain rate on the averaging scale through a description of rain statistics in terms of fractals. An alternative approach that leads to an explicit form of the space-time covariance of the rain rate exhibiting the observed power law scaling behavior is based on a stochastic dynamical equation for the spatial Fourier components of the point rain rate field [*Bell* 1987; *Bell and Kundu* 1996; *Kundu and Bell* 2003, 2006]. The model depicts the rain field as undergoing anomalous or fractional diffusion driven by white noise and incorporates in a natural way the observed dependence of the correlation time scale on the degree of spatial averaging. The model leads to prediction of dynamical scaling [*Kundu and Bell* 2006], a form of invariance of the statistics under a combined space-time scale transformation observed in isolated storms by *Venugopal et al.* [1999]. A major limitation of the latter type of model is that unlike the fractal models it is restricted to describing only the second moment statistics.

In the present work we restrict ourselves to studying the spatial statistics of rain from a phenomenological point of view leaving aside the temporal dependence aspects. We are concerned with the spatial statistics of instantaneous area-averaged rain rate $r_L$ obtained by averaging over an $L \times L$ square. Assuming space-time homogeneity of the statistics, the entire gamut of statistical properties of the non-negative random variable (RV) $r_L$ can be completely derived from the probability density function (pdf) $f(r_L; c_i(L))$ where $c_i(L)$ ($i = 1, 2, \ldots$) are a set parameters depending on the averaging length scale $L$. In this paper we propose a new candidate for the pdf of $r_L$ based on an analysis of a gridded data set of surface radar measurements of rainfall. Our study is based on a gridded precipitation data set [*Short et al.* 1997] constructed from the radar scans obtained during the Tropical Ocean Global Atmosphere -- Coupled Ocean Atmosphere Response Experiment (TOGA-COARE) [*Webster and Lukas* 1992].

Spatial intermittence of rain implies that $r_L$ has a mixed distribution with a non-zero probability $\Pr[r_L = 0] \equiv 1 - p(L)$ of attaining the sharp value 0, where $p(L) = \Pr[r_L > 0]$ represents the probability that an $L \times L$ grid box contains non-zero rain. Clearly $p(L)$ must be an increasing function of $L$. If a rain image is "coarse-grained" through a scale transformation $L \rightarrow L' = \lambda L$, ($\lambda > 1$), adjacent rainy and non-rainy patches are subsumed within a larger area which is designated as rainy; consequently $p(\lambda L) > p(L)$ ($\lambda > 1$). Conversely, magnifying a rain image to a finer spatial resolution $L' < L$, in general reveals rainy and non-rainy areas interspersed within a rainy area at resolution $L$. Indeed it seems possible to assume that $p(L) \rightarrow 0$ as $L \rightarrow 0$. Radar observations of rain are often well described by a power law dependence on $L$: $p(L) \sim L^{\chi}$ ($\chi > 0$) for small $L$, suggesting an underlying fractal structure of the support of the rain field of fractal dimension $\chi$ [e.g. *Kedem and Chiu* 1987a, *Over and Gupta* 1994, *Kundu and Bell* 2003]. The exponent $\chi$ represents the intermittency exponent for the spatial rain field. A class of cascade models, known as the $\beta$-model in the context of turbulence theory [*Frisch et al.* 1978], with a finite probability of zero generator yields a near power law dependence of $p(L)$ on $L$ [*Over and Gupta* 1994]. However, the power law behavior must break down at large scales since $p(L) \leq 1$ for all $L$ and is expected to approach unity as $L \to \infty$. As noted by *Kedem and Chiu* [1987a], the simple fact that $p(L)$ depends on $L$ already precludes the area-averaged rain rate field from being self-similar.

A lognormal distribution has often been used to describe the continuous part of the distribution corresponding to $r_L > 0$ [*Biondini* 1976, *Lopez* 1977, *Houze and Cheng* 1977, *Crane* 1986, *Kedem and Chiu* 1987b], which is empirically known to be unimodal and highly skewed to the right with a rapidly decaying tail as $r_L \to \infty$. From a modeling perspective, lognormal distribution is attractive, since it naturally arises in a multiplicative process involving independent identically distributed (iid) RVs with a finite mean and variance because of the Central Limit Theorem [*Feller* 1971]. A lognormal multiplicative cascade model of energy transfer across eddy size scales was originally invoked to account for intermittency in fully developed turbulence [*Kolmogorov* 1962, *Obukhov* 1962]. However, there is at present no consensus on the "correct" pdf for rain rates. A number of authors have found evidence of departure from the lognormal distribution [e.g. *Martin* 1989, *Pavlopoulos and Kedem* 1992, *Kedem et al.* 1994, *Jameson and Kostinski* 1999]. Several other distributions, including the gamma [*Ison et al.* 1971] and Weibull [*Wilks* 1989], have also been used to represent statistics of precipitation data. For a recent comparative study of the lognormal and gamma distributions, see *Cho et al.* [2004]. Scale dependence of the spatial gradient of rain rate was explored by *Kumar and Foufoula-Georgiou* [1993] and *Perica and Foufoula-Georgiou* [1996] for individual storm events using wavelet transform method, who attempted to relate the statistical parameters to physical storm characteristics. Our primary interest is in a statistical description of rainfall climatology over a relatively large area and a long period of time, which includes the totality of all rain events as well as the non-rainy regions in a certain space-time volume. We find that the lognormal distribution is unsatisfactory at coarser spatial resolutions. This leads us to seek a new probability distribution that better represents rainfall statistics over a broad range of spatial scales. We confine our search within the class of the so-called *infinitely divisible* (ID) *distributions* for candidates suitable for describing the distribution of $\ln r_L$. They emerge as probability distributions of the sum of an arbitrary number of iid RVs and contain many distributions commonly used in hydrology, including the normal (and more generally, the Lévy stable), Poisson, gamma and Gumbel extreme value distributions as special cases. They also naturally arise in the context of a multiplicative cascade process, which involves exponentiation of additive iid RVs.

In our search for a suitable probability distribution describing the spatial statistics of $r_L$ from a precipitation data set, we depart from the usual approach in which the parameters of an empirically chosen form of the pdf are determined by directly fitting it to the rain rate histograms. Instead, we develop a theoretical method in which the pdf is constructed from the moments of the data by utilizing some mathematical characterizations of ID distributions. A set of auxiliary dimensionless quantities constructed from the moments serves to guide us in identifying a suitable member of this class and the parameters of the distribution are estimated by fitting these functions to data. The family of distributions that we construct contains a certain type of Lévy-stable distribution as limiting case.

An important feature of the rain rate field is that the moments of the rain intensity field conditional on nonzero rain $m(q;L) = \langle r_L^q | r_L > 0 \rangle$ (where angle brackets $\langle \ldots \rangle$ denote ensemble average with respect to the distribution of $r_L$ conditioned on $r_L > 0$) exhibit power law scaling with respect to $L$: $m(q;L) \propto L^{-\eta(q)}$. In general the scaling exponents $\eta(q)$ have a nonlinear dependence on the moment order $q$ – a property commonly referred to as multifractal or multiscaling behavior. In the case of fully developed turbulence, the relevant quantities are the statistical moments of the (longitudinal) velocity difference across a distance $L$, $\langle (\delta v_L)^q \rangle$, and the spatially averaged energy dissipation $\langle \varepsilon_L^q \rangle$. Various phenomenological models have been used with considerable success to account for their scaling properties [*Kolmogorov* 1962, *Benzi et al.* 1984, *Parisi and Frisch* 1985, *Meneveau and Srineevasan* 1987a,b, *She and Leveque* 1994, *Dubrulle* 1994, *She and Waymire* 1995]. They lead to simple explicit formulae for the corresponding scaling exponents as function of $q$ that often agree remarkably well with observation. See *Frisch* [1995] for a detailed account of multiscaling properties of fully developed turbulence. *Novikov* [1994] has explored the possibility of understanding the scaling behavior of the turbulent energy dissipation field in terms of an underlying log-ID distribution. Power law scaling of precipitation statistics has been known empirically for some time and significant efforts have been made to explain it in terms of multiplicative cascade processes [*Schertzer and Lovejoy* 1987, *Gupta and Waymire* 1990, 1993]. While atmospheric turbulence certainly plays a role in determining the space-time distribution of precipitation, rainfall microphysics introduces additional complexity into the precipitation process and there is no compelling reason why the exponents resulting from turbulence models would as such carry over to rain statistics. From the dependence of the parameters of the probability distribution on the spatial averaging length $L$ we will be able to estimate the scaling exponents approximately and compare the predicted values with those estimated directly from the moments of the data.

The paper is organized as follows. In Sec. 2 we provide a brief review of the basic definitions and some of the mathematical properties of the ID distributions that are needed in our investigation. We then describe the mathematical method employed in this paper to compute the pdf of a log-ID distribution from certain auxiliary quantities constructed from the moments of the distribution. In Sec. 3 we give an account of the analysis of the TOGA-COARE data set. Sec. 4 is devoted to the proposed new probability distribution with the necessary mathematical characterizations for comparison with data. In Sec. 5 we present the results in detail and discuss the various outstanding issues. Sec. 6 summarizes the main conclusions and suggests some directions for future work. A number of mathematical details relevant to our work are relegated to three appendices so as to avoid distraction. They include derivations of some results used in the main text and can be skipped on a first reading if desired.

## 2.  Theoretical Preliminaries.

The notion of infinite divisibility naturally arises within the theory of probability distributions of a sum of an arbitrary number of iid RVs in connection with the familiar Central Limit Theorem.  The concept is originally due to *de Finetti* [1929], with further seminal contributions by Kolmogorov, Lévy and Khintchine. Detailed mathematical expositions of the subject can be found in a review article by *Bose et al* [2002] and in the monographs by *Feller* [1971], *Lukacs* [1970] and *Steutel and van Harn* [2004]. The last one is also a very complete source of references to the original papers.

### 2.1 ID Distributions.

In this subsection we present a number of mathematical results regarding a subclass of ID distributions that arise in our investigation.

A RV $X$ is said to be ID if for *every* positive integer $n$, $X$ can be expressed (in distribution) as the sum of $n$ iid RVs $X_{n,j}$ ($j = 1, 2, …, n$). The characteristic function (CF) $\phi(t) \equiv E[e^{itX}] = \int_{-\infty}^{\infty} g(x)e^{itx}dx$ is the $n$-th power of $\phi_n(t)$ ($E[…]$ denotes expectation value), the CF of $X_{n,j}$, i.e. $\phi(t) = [\phi_n(t)]^n$. The pdf of $X$, namely $g(x)$ (which is given by the inverse Fourier transform of $\phi(t)$) is the $n$-fold convolution of $g_n(x)$, the pdf of $X_{n,j}$; symbolically, $g(x) = [g_n(x)]^{*n}$. In the context of rainfall statistics $X$ will represent a logarithmic rain rate variable to be introduced later.

A random process consisting of stationary independent increments naturally leads to an ID distribution. Consider a sequence of RVs $Y(\lambda)$ parameterized by a real variable $\lambda$ such that the difference $Y(\lambda+\lambda_o) - Y(\lambda_o) = X(\lambda)$  is function of $\lambda$ alone and the increments $X_{n,k} = Y(\lambda_k) - Y(\lambda_{k-1})$ ($k = 1, 2, …, n$) are iid RVs, where $\lambda_o, \lambda_1, …, \lambda_{n-1}, \lambda_n = \lambda+\lambda_o$ are a set of ($n$+1) equally spaced points separated by $\lambda/n$. Since each $X_{n,k}$ is distributed like $X(\lambda/n)$ and the choice of $n$ is completely arbitrary, it follows that $X(\lambda)$ is ID.

Next we state a fundamental result giving a necessary and sufficient condition that a function $\phi(t)$ is the CF of an ID distribution [see e.g. *Lukacs* 1970, *Steutel and van Harn* 2004]:

**The Lévy canonical representation** – A complex-valued function $\phi(t)$ of a real variable $t$ is the CF of an ID distribution iff  it can be expressed in the form

$$\phi(t) = \exp\left[i\kappa t - \tfrac{1}{2}\sigma^2 t^2 + \int_{\mathbf{R}\setminus\{0\}}\left(e^{itu} - 1 - \frac{itu}{1+u^2}\right)dH(u)\right], \tag{2.1}$$

where $\kappa$ is real, $\sigma^2$ is real and $\geq 0$, and $H(u)$ is a non-decreasing right-continuous function (the Lévy spectral function) on $(-\infty,0)$ and $(0,\infty)$, so that  $H(u) \to 0$ as $u \to \pm\infty$ and $\int_{[-\varepsilon,\varepsilon]\setminus\{0\}} u^2 dH(u)$ is finite for every $\varepsilon > 0$. The representation is unique.

Eq. (2.1) allows one to represent such an RV, up to a shift, as a sum of a Gaussian RV and a limiting sum of (suitably scaled) independent Poisson distributed RVs.

The four parameter Lévy stable distributions $S_\alpha(c,\beta,\kappa)$, where $\alpha$ $(0<\alpha<2)$ is the stability index, and $\beta$ $(-1\leq\beta\leq1)$, $c$ $(0<c<\infty)$ and $\kappa$ $(-\infty<\kappa<\infty)$ are, respectively, the asymmetry, scale and location parameters, constitute an important subclass of ID distributions. The $\alpha=1$ Lévy stable distributions are distinguished from the others by their unusual transformation property under a rescaling of variables and have to be treated separately. The maximally asymmetric distribution $S_1(c,-1,0)$ turns out to play an important role in the course of our investigation. It has the CF

$$\phi(t) = \exp\left[-c|t| + (2i/\pi)ct\ln|t|\right]$$

See Appendix A for a short review and *Samorodnitsky and Taqqu* [1994] for further details of various properties of the stable distributions.

From the CF $\phi(t)$ one can construct a function $a(q) = E[e^{qX}]$ of a real variable $q$ via analytic continuation $it \to q$, assuming that one does not encounter singularities in the complex $t$-plane. The quantity $a(q)$ has the mathematical interpretation of being the $q$-th order moment of the RV $Y = e^X$. Note that $a(q)$ can also be regarded as the moment generating function for $X$. The Taylor expansion of $a(q)$ at $q=0$, if it exists, yields the successive integer order moments of $X$. If the distribution of $X$ is ID, one can easily derive an integral representation of $a(q)$ from the Lévy canonical representation (2.1). For the purpose of our intended application to the rainfall problem, guided by hindsight (see discussion at the end of Sec. 5), we limit ourselves to the special case

$$\sigma^2 = 0, \ \ H(u>0) = 0. \tag{2.2}$$

Letting $u\to-u$ and introducing $h(u) = H(-u)$ when $u>0$ we obtain the integral representation

$$\ln a(q) = q\kappa + \int_{0+}^{\infty}\left(1 - e^{-qu} - \frac{qu}{1+u^2}\right)dh(u). \tag{2.3}$$

The maximally asymmetric Lévy stable distribution $S_1(c,-1,0)$ corresponds to the choice $h(u) = (2c/\pi u)$ $(u>0)$ and in this case for $q>0$, $a(q)$ is given by the simple formula [*Gupta and Waymire* 1990, *Samorodnitsky and Taqqu* 1994]

$$\ln a(q) = (2c/\pi)q\ln q \tag{2.4}$$

Note that it has a singularity at $q=0$, consistent with the fact that the moments of $X$ do not exist for this distribution. The distribution is strongly skewed to the left; it has a power law tail at large negative $X$

$$\Pr\left[X<-x\right] \sim (2c/\pi)x^{-1}, \ \ x\to\infty \tag{2.5}$$

but falls off steeply at large positive $X$ [*Samorodnitsky and Taqqu* 1994]

$$\Pr\left[X > x\right] \sim \frac{1}{\sqrt{2\pi}} \exp\left[-\frac{(\pi x/2c) - 1}{2} - e^{(\pi x/2c) - 1}\right], \quad x \to \infty. \tag{2.6}$$

## 2.2 Application to rainfall statistics.

The results summarized above are now applied to the problem of interest, namely quest for the distribution of the rain rate variable $r_L$ within the class of log-ID distributions. Since the rain rate is governed by a mixed distribution, the normalization of the pdf describing the continuous part of the distribution can be expressed as

$$\int_0^\infty dr_L \, f\left(r_L; c_i(L)\right) = p(L). \tag{2.7}$$

where $p(L) = \Pr[r_L > 0]$. We define

$$\mu(q;L) = \left\langle r_L^q \right\rangle = \int_0^\infty dr_L \, r_L^q f\left(r_L; c_i(L)\right) \tag{2.8}$$

as a function of the moment order $q$ where $q$ is a real variable. When $q > 0$, $\mu(q;L)$ are to be interpreted as the unconditional moments of $r_L$. Moreover, $p(L) = \lim_{q \to 0} \mu(q;L)$. The moments conditional on non-zero rain are given by

$$m(q;L) \equiv \left\langle r_L^q \mid r_L > 0 \right\rangle = \mu(q;L)/p(L) \tag{2.9}$$

and are meaningful for all $q$, both positive and negative. For convenience, we introduce the "dimensionless moments"

$$a(q;L) = m(q;L)/\left[m(1;L)\right]^q. \tag{2.10}$$

Note that by definition $a(0;L) = a(1;L) = 1$. When expressed in terms of the dimensionless logarithmic rain rate variable

$$x_L = \ln\left[r_L/m(1;L)\right] = \ln\left[p(L) \, r_L/\langle r_L\rangle\right], \tag{2.11}$$

the function $a(q;L)$ has the simple interpretation

$$a(q;L) = E\left[\exp(q x_L)\right] \equiv \sum_{n=0}^\infty \frac{q^n}{n!} E\left[x_L^n\right], \tag{2.12a}$$

i.e. $a(q;L)$ is the moment generating function of the RV $x_L$, (the condition $r_L > 0$ being automatically satisfied in the expectation value $E[\ldots]$ computed with respect to the distribution of $x_L$). In classical probability theory the coefficients of the Taylor series expansion of $\ln a(q;L)$ at $q = 0$ define the successive cumulants of $x_L$:

$$\ln a(q;L) = \sum_{n=1}^\infty \frac{\kappa_n(L)}{n!} q^n \tag{2.12b}$$

This cumulant expansion later plays a central role in our analysis. The pdf of the variable $x_L$, $g(x_L;L)$ (which has $L$-dependence through $x_L$ as well as through the parameters of the distribution) is given by the expression

$$g(x_L;L) = r_L f(r_L;c_i(L))/p(L). \qquad (2.13)$$

It is the inverse Fourier transform of the CF $\phi(t;L) = E[\exp(itx_L)]$, i.e.

$$g(x_L;L) = \frac{1}{2\pi} \int_{-\infty}^{\infty} dt e^{-itx_L} \phi(t;L) \qquad (2.14)$$

and is normalized as

$$\int_{-\infty}^{\infty} dx_L \, g(x_L;L) = 1. \qquad (2.15)$$

Since the pdf of a distribution is necessarily nonnegative, it follows from a classical result [Bochner's theorem; see e.g. *Lukacs* 1970] that $\phi(t;L)$ is a positive definite function, i.e. satisfies the inequality (bar denotes complex conjugation)

$$\sum_{k=1}^{n} \sum_{j=1}^{n} \phi\left(t_k - t_j;L\right) \bar{\xi}_k \xi_j \geq 0$$

for all $n \geq 1$, real $t_j$, $t_k$ and complex $\xi_j$, $\xi_k$ ($j,k = 1,2, \ldots, n$).

Ideally one would like to construct the pdf $g(x_L;L)$ directly from the dimensionless moment function $a(q;L)$ determined from the data for all $q$. However attempts to describe the moment data in terms of simple empirical functions in general yield a non-positive definite function $\phi(t;L)$ thus leading to potential candidates for the pdf that become negative in certain ranges of $x_L$ and are therefore mathematically unacceptable. In order to overcome this obstacle we restrict ourselves to the family of ID distributions for which an explicit representation of $a(q;L)$ can be constructed.

In particular, we wish to consider the case when the distribution of $x_L$ belongs to the subclass satisfying Eq. (2.2) and characterized by a Lévy spectral function $h(u;L)$ with the following properties: (i) $h(u;L)$ is a non-increasing left-continuous function on $(0,\infty)$, (ii) $h(u;L) \to 0$ faster than exponential as $u \to \infty$ and (iii) the integral $\int_0^\varepsilon u^2 dh(u;L)$ is finite for every $\varepsilon > 0$. The conditions (i) and (iii) are the same as in Eq. (2.1), while the more stringent fall-off behavior of $h(u;L)$ in condition (ii) ensures the convergence of the integral representation given by Eq. (2.3) for all $q$. Under these conditions it is possible to re-express the function $\ln a(q;L)$ in the simpler form (see Appendix B for derivation)

$$\ln a(q;L) = q \int_0^\infty du \, h(u;L)\left[e^{-u} - e^{-qu}\right]. \qquad (2.16)$$

The problem of finding a suitable distribution within this subfamily thus reduces to finding a suitable function $h(u;L)$ so that the theoretically computed $a(q;L)$ agree with those estimated from data.

For a lognormally distributed $r_L$ (i.e. normally distributed $x_L$), one has the simple form [*Aitchison and Brown* 1957] with a parabolic $q$-dependence,

$$\ln a(q;L) = \tfrac{1}{2}\sigma^2(L)\left(q^2 - q\right), \qquad (2.17)$$

where $\sigma^2(L)$ is the variance of $x_L$. It arises as the special case $\sigma^2 \neq 0$, $H(u) = 0$ in Eq. (2.1) and will turn out *not* to describe the large order moments of radar precipitation data well.

# 3. Data Analysis.

Our analysis utilizes a gridded precipitation data set described in [*Short et al.* 1997]. The data set was constructed from radar scans that were collected during TOGA-COARE, an experimental campaign conducted in the tropical western Pacific during the period November 1992 to February 1993 using two ship-borne Doppler radars (labeled TOGA and MIT). The entire data set consists of 101 days of observation divided into three approximately month-long "cruises" in which radar images were available about every ten minutes. In this paper we present results based on the data obtained during Cruise 3, which contains 4380 merged rain images from both radars. Each rain image consists of a $278 \times 278$ array of pixels 2 km $\times$ 2 km in size. The statistics were collected from $128 \times 128$ km$^2$ areas concentric with the circular radar fields of view, as described by *Kundu and Bell* [2003]. Statistics for all $L \times L$ sub-areas with $L = 2, 4, 8,$ …,128 km were computed by aggregating the $L = 2$ km single pixel data. Only those grid boxes in a rain map were used for which at least 95% of the box had valid data. This was done in order to exclude boxes, especially those at the smaller scales 4, 8, and 16 km located near the center, which occasionally suffered from data dropout. The algorithm for moment computation was formulated in double precision arithmetic and was tested to ensure that the results are not compromised by machine round-off error.

We note that the rain rate variable $r_L$ is to be interpreted as the spatial average of an underlying instantaneous point rain rate $r(\mathbf{x},t)$ (which itself is not directly measurable) over an $L \times L$ grid box, i.e. $r_L \equiv r_L(t) = L^{-2}\int_{L\times L} r(\mathbf{x},t)d^2\mathbf{x}$. The mean rain rate $\langle r_L \rangle \equiv \mu(1;L)$ would in general not be independent of $L$ (and time) unless appropriate spatio-temporal homogeneity assumptions are made regarding the statistics of the random variable $r(\mathbf{x},t)$ over the entire area of interest and the entire period of observation. However, given a dataset consisting of single pixel data at a fixed resolution $L = L^*$ (= 2 km) which defines a certain minimum scale, our coarse-graining procedure automatically enforces the condition $\langle r_L \rangle = \langle r_{L*} \rangle \equiv \langle r \rangle$ at each explored spatial scale $L \geq L^*$ up to the largest scale $L = L_0$ (= 128 km). The quantity $\langle r \rangle$ can then be interpreted as an estimate of the (scale-independent) mean of the probability distribution we seek for describing the rain rate statistics over the entire $L_0 \times L_0$ area and for the entire observation period $0 \leq t \leq T$ under the presumed homogeneity conditions. Ideally, one should also test the data for statistical homogeneity — a somewhat arduous task that we have refrained from carrying out in detail.

For the MIT radar, about 78.3% of the pixels had valid data and of the latter about 10.6% had non-zero rain. For the TOGA radar the corresponding fractions are 76.0% and 8.4% respectively. Most of the missing data was from images taken when one of the radars was not operational. At each spatial scale $L$, the quantities $p(L)$ and the moments $m(q;L)$, and from them the dimensionless moments $a(q;L)$ were evaluated for various values of $q$ between $-2$ and 10. Also the appropriately normalized rain rate histograms were computed for each $L$ in equal intervals of

$$x_L = \ln[p(L)r_L/\langle r \rangle]. \tag{3.1}$$

For MIT and TOGA Cruise 3, the mean rain rate was estimated to be $\langle r \rangle = 0.200 \pm 0.036$ and $0.155 \pm 0.035$ mm h$^{-1}$ respectively. The standard error estimates are obtained under the assumption of (asymptotic) normality of the sample mean (by virtue of the Central Limit Theorem, even though the individual $r_L$ are markedly non-gaussian). Although the successive radar scans are about 10 min. apart and as a result there are a large number of images available ($N = 4380$), the values of $r_L$ for $L = 128$ km are strongly time-correlated and therefore cannot be treated as independent samples. We assume an exponential time autocorrelation typifying red noise in order to make use of an error estimate obtained by *Leith* [1973]. *Leith*'s estimate can be expressed in the form $s_L/\sqrt{N_{\text{eff}}}$, where $s_L^2$ is the variance of $r_L$, $N_{\text{eff}} = T/(2\tau_L)$ is the effective number of independent samples in the time series of length $T \approx 30$ days and $\tau_L$ is the $(1/e)$-folding autocorrelation time of $r_L$. For MIT and TOGA Cruise 3, *Kundu and Bell* [2003] found the values $s_L^2 = 0.17$ and 0.12 mm$^2$ h$^{-2}$, $\tau_L = 2.8$ and 3.6 h, which yield $N_{\text{eff}} \approx 129$ and 100 respectively. An obvious caveat in this computation is that contrary to the assumption made in *Leith*'s [1973] original derivation, the observed lagged autocorrelation functions are in general markedly non-exponential. Nonetheless, we believe that the error estimates have the right order of magnitude.

## 4. The Proposed Distribution.

In order to search for an appropriate candidate for the rain rate distribution, we examine the $q$-dependence of the dimensionless moments $a(q;L)$ for each $L$ ("the moment curves", Fig.1). It is found convenient to carry this out in terms of the auxiliary variable

$$\Lambda(q;L) = q^{-1}\ln\ a(q;L) \tag{4.1}$$

We find that when $q > 1$, $a(q;L)$ closely obeys the formula

$$\ln\ a(q;L) \approx (2/\pi)c(L)q\ \ln q\ , \tag{4.2}$$

where $c(L) > 0$, characterizing a log-$S_1(c,-1,0)$ distribution [see Eq. (2.4)], but crosses over to a different $q$-dependence when $q < 1$. Systematic departure from Eq. (4.2) at small $q$ becomes apparent when one examines the plots of $\Lambda(q;L)$ vs. $q$ (Fig.1). In fact it is found that the numerical behavior of $\Lambda(q;L)$ as function of $q$ near $q = 0$ is strongly indicative of continuity of the slope $\Lambda'(q;L) \equiv d\Lambda(q;L) /dq$ at $q = 0$ instead of the $1/q$ divergence predicted by (4.2). If rain rates were lognormally distributed, one would have had, according to Eq. (2.17), $\ln a(q;L) \propto q^2 - q$, and consequently $\Lambda'(q;L) = $ const. for all $q$. This also explains why a nonzero $\sigma^2$ in Eq. (2.1) is not favored by data: any such term would lead to a quadratic dependence on $q$ at large $q$ in Eq. (4.2).

The asymptotic large-$q$ behavior (4.2) implies that the right tail of $g(x_L;L)$ representing the high rain rate events resembles that of the stable distribution $S_1(c,-1,0)$ which drops off precipitously as $x_L \to \infty$ $(r_L \to \infty)$ as indicated by Eq. (2.6). The large-$q$ behavior of $a(q;L)$ is governed by small-$u$ behavior of the Lévy spectral function $h(u;L)$ in the integrand of Eq. (2.16) and vice versa. Since the distribution $S_1(c,-1,0)$ corresponds to the choice $h(u;L) = 2c(L)/\pi u$ in Eq. (2.16), we surmise that an appropriate modification will need to preserve the $u^{-1}$ dependence near the origin. We find that the simple choice

$$h(u;L) = \begin{cases} 2c(L)/\pi u & ; \; u \leq b(L) \\ 0 & ; \; u > b(L) \end{cases}, \qquad (4.3)$$

where $c(L)$ and the cut-off $b(L)$ are scale-dependent parameters, in fact reproduces the essential aspects, if not the details, of the $q$-dependence of the moments including both the large-$q$ behavior given by Eq.(4.2), as well as the behavior near $q = 0$ (Fig. 1).

For the ID distribution defined by the choice (4.3) for $h(u;L)$, the integral (2.16) converges for all $q$ and yields the explicit formula

$$\ln a(q;L) = (2/\pi)c(L)q \left[ \ln|q| + \text{Ei}(-b(L)) - \text{Ei}(-b(L)q) \right], \qquad (4.4)$$

where $\text{Ei}(x) = -\int_{-x}^{\infty} dt (e^t/t)$ denotes the exponential integral function [*Abramowitz and Stegun* 1972]. Eq. (4.4) predicts that, despite its appearance, $\ln a(q;L)$ is in fact analytic at $q = 0$ and can be written for small $q$ as a series expansion (the logarithmic branch point singularities cancel)

$$\ln a(q;L) = q(q-1)[\, c_0(L) + c_1(L)q + c_2(L)q^2 + \; \ldots]. \qquad (4.5)$$

The coefficients $c_0(L)$, $c_1(L)$ etc. are simple linear combinations of the cumulants $\kappa_n(L)$ introduced in Eq. (2.12b): $c_0 = -\kappa_1$, $c_1 = -(\kappa_1 + \frac{1}{2} \kappa_2)$, etc.. Moreover, in view of the series expansion [*Abramowitz and Stegun* 1972]

$$\text{Ei}(-x) = \gamma + \ln x + \sum_{n=1}^{\infty} \frac{(-x)^n}{n \cdot n!} \qquad (x > 0)$$

they can be expressed in terms of $c(L)$ and $b(L)$:

$$c_0 = (2c/\pi)[\gamma + \ln b - \text{Ei}(-b)], \; c_1 = c_0 - (2cb/\pi), \; c_2 = c_1 + (1/2.2!)(2/\pi)cb^2, \qquad (4.6)$$

and so on (suppressing the $L$-dependence for ease of notation), where $\gamma = 0.5772 \ldots$ denotes Euler's constant. We remark that Eq. (4.4) involves an overall compromise, which sacrifices slightly the quality of fit at large $q$ in order to achieve an acceptable fit at small $q$. It generates an additional subdominant term linear in $q$ that survives at large $q$ in contrast to the simpler Eq. (4.2), which by itself fits the large order behavior of the moments quite well. Finally, we note that since the function $a(q;L)$ defined by Eq. (4.4) is an entire function (i.e. analytic in the entire complex $q$-plane), it immediately follows that our postulated probability law for $r_L$ yields finite moments of all orders $q$, both positive and negative.

A limiting case of interest arises when $c \to \infty$, $b \to 0$ in such a way that $cb$ tends to a finite value. In this limit $c_0 = 2cb/\pi$ and the higher order coefficients $c_1$, $c_2$ etc. all vanish as successive powers of $b$. The series expansion (4.5) then reduces to the simple parabolic form (2.17) with $\sigma^2 = 2c_0 = 4cb/\pi$ and consequently the new distribution approaches a lognormal distribution.

For the CF $\phi(t;L)$ we obtain, after carrying out an analytic continuation $q \to it$, the explicit formula

$$\ln \phi(t;L) = (2/\pi)c(L)\left[-|t|\operatorname{Si}(b(L)|t|) + it\{\ln|t| - \operatorname{Ci}(b(L)|t|) + \operatorname{Ei}(-b(L))\}\right], \quad (4.7)$$

where $\operatorname{Si}(z) = \int_0^z dt\,(\sin t/t)$ and $\operatorname{Ci}(z) = -\int_z^\infty dt\,(\cos t/t)$ are the standard sine and cosine integral functions [*Abramowitz and Stegun* 1972].

# 5. Results and Discussion.

Because of the intricate relationship between the moments and the overall shape of the pdf, obtaining a set of parameter values that best describe the data proved to be a somewhat delicate problem. We first obtained a preliminary estimate by leaving $\langle r \rangle$ and $p(L)$ fixed at their sample values and evaluating $c(L)$ and $b(L)$ by a nonlinear least squares fit to the moment curves through Eq. (4.4) in the range $0 \le q \le 10$ using the routines provided by *Press et al* [1995]. The estimates of $c(L)$ and $b(L)$ thus obtained were found to be linear in $\ln L$ to considerable accuracy. They were taken as starting point for constructing the pdf $g(x_L;c(L),b(L))$ from $\phi(t;L)$ by numerically evaluating the inverse Fourier transform for each $L$. We sought to improve the estimates by examining the overall quality of the fit to the rain rate histograms. The final parameter choices obtained by essentially a trial-and-error approach represent our best effort to reproduce both the moment curves and the observed histograms as faithfully as possible. They are listed in Table 1. (See Appendix C for some computational details). We see that the computed pdfs for both MIT Cruise 3 and TOGA Cruise 3 fit the observed rain rate histograms fairly well over the entire explored range of spatial scales (Fig. 2). This is also supported by the quantile plots showing the $k$-th quantile, $x_L*(k)$ $(0 < k < 1)$ satisfying the equation $\Pr[x_L \le x_L*(k)] = k$ computed from sample data against those predicted from the model. Fig. 3 shows these plots for quantiles $k$ in the range $0.001 \le k \le 0.999$. They reinforce the conclusion that the proposed distribution successfully reproduces the observed frequencies of high rain rate events. The model correctly reflects the gradual change in the shape of the distribution with the increase of $L$, as more and more areas of zero rain interspersed among the rainy areas are averaged together. In particular, it accounts for the slowly decaying tail at lower rain rates that becomes increasingly pronounced at larger $L$, an effect not explained by a lognormal distribution. Fall-off of the pdf at high rain rates governed by the tail behavior (2.6) is much more rapid than what would be expected from a lognormal distribution. It can be argued that the observed pdfs for intermediate $L$ are more representative of the "true" statistical behavior of precipitation since spatial aggregation serves to smooth out possible data preparation artifacts introduced at the original grid scale. Such artifacts can arise, for example, from radar algorithm error due to misclassification of rainy pixels into convective/stratiform types and from uncertainties in detecting low rain rates at the intrinsic spatial resolution of the radar. On the other hand, as $L$ becomes large and comparable to the synoptic scale, one can expect effects of spatial inhomogeneity to distort the results. We also notice a systematic discrepancy between the data and the model at low rain rates that prominently appears in the quantile plots in Fig. 3. We find that as $k \to 0$, the data quantiles rapidly flatten out to a constant value of $x_L$ corresponding to a minimum nonzero rain rate $r_L$ of about 0.01 mm h$^{-1}$ at the pixel scale $L = 2$ km that is registered by the radar whenever it detects rain. We interpret this as a low rain sensitivity threshold for the radar possibly arising from its intrinsic electrical noise. The discrepancy between the data and the model may therefore be due, at least in part, to the radar measurements of low rain rates being inherently unreliable and possibly noise-limited. The difficulty of detecting low rain rates may also affect identification of the support of the nonzero rain field, i.e. estimates of the parameter $p(L)$. The other TOGA-COARE data subsets also exhibit similar general behavior with only minor individual differences, except for TOGA Cruise 1 in which a second weak mode seems to appear in $g(x_L)$ at very low rain rates as $L$ is increased.

Fig. 4 depicts the scale dependence of the model parameters $\ln p(L)$, $c(L)$ and $b(L)$ regressed against $\ln L$. Fit to a power law $p(L) \propto L^\chi$ yields an estimate of the intermittency exponent $\chi$: $\chi \approx 0.531$ for MIT Cruise 3 and $\chi \approx 0.546$ for TOGA Cruise 3. From Eq. (4.2) upon setting $q = 2$ it follows that

$$c(L) \approx \frac{\pi}{4} \log_2\left[ \frac{p(L)\mu(2;L)}{\langle r \rangle^2} \right] = \frac{\pi}{4} \log_2\left[ 1 + \zeta^2(L) \right], \qquad (5.1)$$

where $\zeta(L)$ denotes the coefficient of variation conditional on $r_L > 0$. Based on radar and rain-gauge data sets it has been suggested that $\zeta(L)$ is independent of $L$ [*Short et al.* 1993]. Our results show that this is not true in the present data set. As $L \to \infty$, one expects that $p(L) \to 1$, $\mu(2;L) \to \langle r \rangle$, so that $\zeta(L)$, $c(L) \to 0$. This is consistent with the trend seen in Fig. 4. However, it should be emphasized that the linear dependence on $\ln L$ implied in the regression can only be valid in a limited range of scales since the model parameters are restricted by the inequalities $p(L) \leq 1$, $c(L) > 0$, $b(L) > 0$.

In order to explore the scaling behavior of the moments we examine plots of $\ln m(q;L)$ vs. $\ln L$ for $L$ between 2 and 128 km and $q$ between $-2$ and 10 computed from data (Fig. 5). The linear regressions reveal an approximate power law relationship

$$m(q;L) \sim L^{-\eta(q)} \tag{5.2}$$

in this range of spatial scales yielding a set of "mean" scaling exponents $\eta(q)$. By definition [Eq. (2.9)] $\eta(0) = 0$ and since the unconditional mean $\mu(1) \equiv \langle r \rangle$ is independent of $L$, it follows that $\eta(1) = \chi$, the intermittency exponent. An exact power law scaling of all the moments with computable exponents would automatically follow from our probability model if $\ln p(L)$ and $c(L)$ are linear functions of $\ln L$ and $b(L)$ is a constant. This is however not actually the case since all three of our model parameters that fit the data exhibit nontrivial $L$-dependence; $\ln p(L)$, $c(L)$ and $b(L)$ are all found to be roughly linear in $\ln L$ (Fig. 4). Nevertheless we still find that, for a wide range of values of $q$, a power law scaling empirically holds to a good approximation, and moreover, the exponents $\eta(q)$ can be fairly accurately determined from just its observed values for the $q = \frac{1}{2}$, 1 and 2 moments. The continuous spectrum of scaling exponents determined from the model therefore describes the multiscaling characteristics of the rain field. However, a more careful analysis shows $\ln m(q;L)$ to be nonlinear in $\ln L$ implying that the exponents defined by Eq. (5.2) are actually slightly $L$–dependent.

To understand the origin of the observed approximate power law scaling we represent $\ln m(q;L)$ and $\ln a(q;L)$ in the form of power series in $\ln L$. To the extent the quadratic and higher order terms can be neglected, the coefficient of the $\ln L$ term can be identified as the scaling exponent introduced in Eq. (5.2). (A similar multiscaling analysis based on cumulant expansion of a distribution has also been employed recently by *Venugopal et al.* [2006].) Although in general $\ln a(q;L)$ has an intricate joint dependence on $q$ and $L$, for the purpose of estimating the scaling exponents, it is convenient to employ separate approximations for "large $q$" ($q \geq 1$) and "small $q$" ($|q| \leq 1$) with separable $q$- and $L$-dependence. To this end we replace Eq. (4.4) by the following approximation:

$$\ln a(q;L) \approx \begin{cases} (2/\pi)c(L)q\ln q & ; (q \geq 1) \\ \left(q^2 - q\right)\left[c_0(L) + c_1(L)q\right] & ; (-1 < q < 1) \end{cases}. \tag{5.3}$$

where $c(L)$, $c_0(L)$ and $c_1(L)$ are assumed to be linear in $\ln L$. We then match the exponents at $q = 1$, $\frac{1}{2}$ and 2 with the observed values by joining the derivative $d\eta(q)/dq$ continuously across $q = 1$. A little computation leads to the following simple formula for $\eta(q)$:

$$\eta(q) \approx \begin{cases} \chi q + \alpha q \ln q & ; (q \geq 1) \\ \chi q + \left[(\alpha + \beta) - (2\alpha + \beta)q\right]\left(q - q^2\right) & ; (-1 < q < 1) \end{cases}, \tag{5.4}$$

where $\alpha = [-2\chi + \eta(2)]/(2\ln 2)$ and $\beta = 4[-\chi + 2\eta(\frac{1}{2})]$, with $\alpha \approx 0.079$, $\beta \approx -0.066$ for MIT Cruise 3 and $\alpha \approx 0.178$, $\beta \approx -0.005$ for TOGA Cruise 3. Results from Eq. (5.4) are plotted in Fig. 6 along with the exponents estimated from data (with error bars representing 95% confidence intervals estimated during the least squares fit to Eq. (5.2)) and those predicted from a simple lognormal distribution obeying Eq. (2.17), namely,

$$\eta_{LN}(q) = \chi q + \tfrac{1}{2}\left[-2\chi + \eta(2)\right]\left(q^2 - q\right). \tag{5.5}$$

The agreement between the predicted and the measured exponents is quite striking, especially considering the heuristic nature of the arguments leading to Eq. (5.4). Only the scaling exponents of the moments of order $q \approx 1$ apparently suffer slightly from the artifacts of our approximation. Not unexpectedly, the model estimates of $\eta(q)$ based on the approximation (5.3) deviate increasingly from the observed values as $q$ becomes more and more negative. In contrast, the lognormal model fares rather poorly in predicting the exponents as $q$ becomes large. The constants $c_0(L)$ and $c_1(L)$ estimated from a least squares fit to $\Lambda(q;L)$ vs. $q$ near the origin are given in Table 1 with the estimated standard errors. We find that the assumption of linearity in ln $L$ anticipated in our explanation of power law scaling holds reasonably well in the data. Also included are the values of $c_0(L)$ and $c_1(L)$ computed from the parameters $c(L)$ and $b(L)$. We see that while the two values of the leading coefficient $c_0(L)$ agree well, for $c_1(L)$ the agreement is rather poor. This can be attributed to the fact that the series expansion (4.5) for ln $a(q;L)$ converges rather slowly and the terms beyond the second are not negligible when $|q| \approx 1$.

At this point it is appropriate to recognize some caveats in our analysis. A complete evaluation of the pdf ideally involves knowledge of the function ln $a(q;L)$ for all $q$, both positive and negative. The asymptotic behavior (4.2) of the moments at large positive $q$ appears to be a robust feature at all spatial scales as evidenced by the fact that the predicted pdf correctly captures the steep fall-off of the observed histogram at each $L$. Thus our proposed distribution accurately describes the relative frequency of spatial occurrence of heavy rainfall. However the $q<0$ moments depend increasingly on radar estimates of very low rain rates, and are thus increasingly unreliable. Our analytic continuation method of inferring the form of the characteristic function $\phi(t;L)$ for all real $t$ implicitly involves additional working assumption about its analytic behavior. The specific distribution that we have proposed based on observed dependence of the moments as function of $q$, has the conceptual advantage that the corresponding ln $a(q;L)$ function given by (4.4) is an entire function, i.e. has a convergent Taylor series expansion everywhere in the complex $q$-plane. However, as mentioned above, practical usefulness of the expansion (4.5) is limited by its slow convergence. Nonetheless, analyticity at $q=0$ allows us to extend the function a certain way into the region of negative $q$. We have decided somewhat arbitrarily to limit ourselves to theoretically exploring the moments only up to $q = -1$. As seen from Fig. 1, the actual $\Lambda(q;L)$ computed from data deviates increasingly from its model-predicted form as $q$ becomes negative. Departure of the shape of the moment curve from the model behavior is seen by examining its slope $\Lambda'(q;L)$ which was also estimated from data. For $\Lambda'(q;L)$ the proposed probability distribution predicts the simple form

$$\Lambda'(q;L) = \left(2c(L)/\pi q\right)\left(1 - e^{-b(L)q}\right) \tag{5.6}$$

which increases monotonically as $q$ decreases. For each $L$ the observed shape generally agrees with the above form up to a certain minimum $q$ where $\Lambda'(q;L)$ attains a local maximum and then decreases slightly to reach a fairly constant value. This departure at negative $q$ is consistent with the fact that the pdf appears to systematically underestimate the low rain rate tail of the histograms at all spatial scales. It remains to be seen whether the discrepancies between the observed histograms and the model pdf are statistically significant.

Limited experimental access to the negative $q$ moments implies a limitation of our knowledge of the corresponding scaling exponents. However, in view of the analyticity at $q=0$, there appears to be no drastic change in behavior as one crosses over from positive to negative $q$ regime. The approximation (5.3) employed to estimate the scaling exponents is designed to preserve the leading asymptotic behavior of the exact model expression for $\ln a(q;L)$ [Eq. (4.4)] for large positive $q$ at the expense of altering somewhat the analytic behavior near $q = 0$.

Our method of estimation of $c(L)$ and $b(L)$ can perhaps be improved. The nonlinear least squares algorithm that we used to obtain the initial estimates by fitting Eq. (4.4) led to large error estimates for these parameters, especially $b(L)$, forcing us into the trial-and-error approach. In retrospect, this is not surprising since the log-Lévy-like tail at high rain rate corresponds to the limit in which $b(L)$ tends to infinity but the rest of the pdf requires a much smaller $b(L)$. Broad internal consistency of our estimates is corroborated by the fact that fitting the parameters to the simple form of $\Lambda'(q;L)$ given by Eq.(5.6) yields values not too far from the values obtained from fitting $\Lambda(q;L)$. Also, one should recognize that a best fit of the $\Lambda(q;L)$ function in a least squares sense does not necessarily lead to best overall fit for the pdf. This is because of the highly nonlinear (and presumably non-local) relationship between the two functions; a small range of $q$ effectively controls the shape of most of the pdf curve, the large $|q|$ moments being instrumental in only determining the tails of the distribution. Our estimates of the parameters should nevertheless be close to being optimal as evidenced by the close overall agreement between the computed pdf and the observed rain rate histograms.

The log-ID distribution we have proposed in this paper has finite moments to all orders and was arrived at by attempting to match the sample moments (which of course are always finite) as closely as possible. In view of the complicated relation between the moments and the pdf noted above, it is in principle possible for a different pdf to provide an adequate fit to data even if its (population) moments cease to exist for orders $q$ outside a certain range. This is a relevant issue since in many multifractal scenarios often the moments of the rain rate distribution beyond a certain maximum and/or minimum order diverge due to the presence of "heavy" (i.e. Pareto-like) tails [e.g. *Lovejoy and Mandelbrot* 1985]. In order to explore this possibility (at each spatial scale $L$), we consider a "test" pdf $g_0(x) \sim C\exp(-\lambda|x|)$ $(\lambda > 0)$ that falls off exponentially as $x \rightarrow \pm\infty$. This corresponds to a power law dependence of the conditional rain rate pdf $f_0(r)$: $f_0(r) \sim r^{\lambda-1}$ as $r \rightarrow 0$ and $f_0(r) \sim r^{-(1+\lambda)}$ as $r \rightarrow \infty$. (The cusp at $x = 0$ is unimportant for our arguments, since we are interested only in the asymptotic behavior). For this pdf the moment function $a_0(q)$ [defined by Eq.(2.12)] has the form $a_0(q) \sim (\lambda^2 - q^2)^{-1}$ and consequently only the moments of order $q$ in the range $|q| < \lambda$ converge. On the other hand, for the proposed distribution, Eq. (4.4) predicts that moments of $r$ exist to all orders. This in particular implies the absence of a Pareto-like power law right tail as $r \rightarrow \infty$. In fact, as we have already noted, for large $q > 0$, $a(q)$ grows approximately like $\exp[(2c/\pi)q \ln q]$ [see Eq. (4.2)] characterizing a log-$S_1(c,-1,0)$ distribution. The expression (2.6) for the tail probability $\Pr[X > x]$ of the associated Lévy stable distribution $S_1(c,-1,0)$ then leads to the asymptotic behavior $f(r) \sim (r/m_1)^{-(1-\delta/2)} \times \exp[-(1/e)(r/m_1)^\delta]$, $m_1 = m(1;L)$, $\delta = \pi/2c$ for the pdf of $r$ (retaining only the dominant term), as $r \rightarrow \infty$. In order to investigate whether the rain rate data itself would allow a heavy-tailed distribution, we proceed with an elementary analysis of the tail quantiles of $x$. For the exponential distribution $g_0(x)$ a simple explicit calculation shows that the $k$-th quantile, $x^*(k)$ obeys a linear relationship with $\ln(1-k)$ at the right tail ($k \approx 1$) and with $\ln k$ at the left tail ($k \approx 0$) with slope $(-1/\lambda)$ in each case. Plots of $x^*(k)$ vs. $\ln(1-k)$ from sample $x$-data however do not exhibit any clear linear regime. Moreover, attempts to estimate the exponent $\lambda$ using the standard Hill estimator [*Hill* 1975] yield values of $\lambda$ that are much larger than unity, rapidly increasing as one considers smaller and smaller ranges of $k$ along the right tail. This is consistent with the (approximate) stretched exponential falloff of $f(r)$ predicted by the model at large $r$.

The situation is not quite as clear with regard to the behavior of $f(r)$ at low rain rates. As the moment order $q \to -\infty$, Eq. (4.4) indicates that $a(q)$ remains finite but grows very rapidly, roughly like $\exp[(2c/\pi)\exp(b|q|)]$. This implies that $f(r)$ tends to zero as $r \to 0$ faster than any power law (but more slowly than, say, the lognormal pdf). In the absence of a closed analytic form, this suggestion is confirmed by a numerical exploration of the asymptotic behavior of the pdf $g(x)$ at large negative $x$. We find that as $x \to -\infty$, the left tail of the computed $g(x)$ can be accurately represented by a simple stretched exponential form $g(x) \sim \exp[-\text{const}.|x|^\nu]$. The exponent $\nu$ (which presumably depends on the parameters $c$ and $b$ in an unknown way) is greater than unity at all explored spatial scales, ranging between the values 1.80 ($L = 2$ km) and 1.52 ($L = 128$ km) for MIT Cruise 3. The faster-than-exponential decay of our model pdf $g(x)$ is clearly consistent with finiteness of the moment function $a(q)$. Over the limited range of $x$ that is experimentally accessible, $g(x)$ is practically indistinguishable from the exponential decay of the test pdf $g_0(x)$ especially for larger $L$ and both fit the left tail of the sample histograms. But examination of the plots of the quantiles $x^*(k)$ from sample $x$-data against $\ln k$ for small $k$ does not reveal a linear regime at any $L$. As we have already noted earlier in this section, some of the disagreement between the data and the theoretical model may be attributable to the inherent difficulty of making accurate radar measurements of low rain rates. The empirical evidence for our pdf is thus somewhat less compelling in the low rain rate regime but we see no evidence of a power law tail at low rain rates in the available data.

A final point to be noted is a rationale for our restriction to the subclass of ID distributions characterized by Eq. (2.2). We have already seen that the large-$q$ behavior of $a(q;L)$ strongly favors $\sigma^2 = 0$. The condition $H(u > 0) = 0$ can conceivably be relaxed if warranted by experimental data to include functions $H(u)$ that fall off sufficiently fast at large positive $u$ (faster than exponential) so that the contribution of the $u > 0$ portion of $H(u)$ to the integral representation of $a(q;L)$ converges for large positive $q$ to ensure the existence of the moments.

# 6. Conclusion.

To conclude, we have introduced a new probability distribution belonging to the log-ID class that describes the spatial statistics of area-averaged rain rate over a broad range of length scales. In view of the Lévy representation (2.1), such a distribution can be interpreted as coming from a multiplicative random process that can be represented as a limiting product of log-Poisson processes. Clarifying the physical significance, if any, of such a representation is an intriguing problem that deserves further investigation.

The scale dependence of the theoretically computed moments of the fitted distribution explains the observed multiscaling of the rain rate field. This allows one to extrapolate rainfall statistics down to sub-grid scales for hydrological applications in a clearly defined and natural manner. Our method of constructing the distribution relies on moment estimation instead of following the traditional route of fitting an empirically chosen form of the pdf directly to the observed rain-rate histograms. Our choice of the Lévy spectral function $h(u)$ should be regarded only as a first guess. Clearly a more systematic method of exploring the entire family of distributions specified by this function will be desirable. It is conceivable that a judicious choice of $h(u)$ will lead to a closer fit for the pdf, perhaps at the cost of introducing more adjustable parameters. Since the pdf is not available in closed analytic form, formulating a systematic tractable method of parameter estimation remains an open problem.

It will also be of interest to explore whether the new distribution can successfully describe the scale dependence of statistics of time-averaged precipitation data derived from rain gauge measurements, which probe the temporal statistics at a point. A dense rain gauge network monitored over an extended period of time provides a natural way to study the statistics of time-averaged rain rate and preliminary investigations with such data appear to be encouraging. We hope to return to this problem elsewhere.

### Acknowledgements

We are grateful to Tom Bell, Anindya Roy and Bimal Sinha for their constant encouragement and many stimulating conversations throughout the course of this investigation. We also thank Tom Bell and Glen Engel-Cox for critically reading the manuscript and suggesting various improvements in the presentation. Constructive comments from two anonymous reviewers greatly helped clarify and enhance several aspects of the analysis. This research was supported by a NASA grant under the Precipitation Measurement Missions (PMM) program.

## Appendix A. Lévy Stable Distributions

As mentioned in the main text, the Lévy stable distribution with $\alpha = 1$, $\beta = -1$ constitute a limiting case of the new distribution. In this Appendix we summarize, for the readers' convenience, some useful properties of the Lévy stable distributions.

The Lévy stable distributions $S_\alpha(c, \beta, \kappa)$ are obtained as a special case of (2.1) in which $\sigma^2 = 0$ and $H(u)$ has the form

$$H(u) = \begin{cases} C_1 |u|^{-\alpha} & ; u < 0 \\ -C_2 u^{-\alpha} & ; u > 0 \end{cases} \tag{A.1}$$

where the constants $C_1$, $C_2$ satisfy $C_1$, $C_2 \geq 0$, $C_1 + C_2 > 0$. They are related to the scale and asymmetry parameters $c$ and $\beta$ through the formulas [*Lukacs* 1970]

$$c^{\alpha} = \gamma(\alpha)(C_1 + C_2) \ , \ \beta = \frac{C_2 - C_1}{C_1 + C_2}. \tag{A.2}$$

The factor $\gamma(\alpha)$ in general has somewhat complicated dependence on the exponent $\alpha$ when $\alpha \neq 1$ and has the value $\gamma(1) = \pi/2$. The CF of the family of distributions $S_{\alpha}(c, \beta, \kappa)$ has the form

$$\ln \phi(t) = \begin{cases} i\kappa t - c^{\alpha}|t|^{\alpha}\left[1 - i\beta \, \mathrm{sgn}(t) \, \tan(\pi\alpha/2)\right] & ; \ \alpha \neq 1 \\ i\kappa t - c|t|\left[1 + i\beta(2/\pi) \, \mathrm{sgn}(t) \, \ln|t|\right] & ; \ \alpha = 1 \end{cases} \tag{A.3}$$

The special case $\sigma^2 \neq 0$, $H(u) = 0$ represents the familiar normal distribution, which can also be regarded as the $\alpha = 2$ limiting case of the Lévy stable distributions (the parameter $\beta$ is redundant in this case). The $\alpha = 1$ stable distributions are distinguished from the others by the fact that multiplication of the random variable by a constant results in rescaling of the scale parameter $c$ accompanied by a nonlinear change of the shift parameter $\kappa$ and have to be treated separately. $S_1(c,0,\kappa)$ denotes the familiar Cauchy distribution. The $\beta = -1$ case is of particular interest to us and corresponds to the choice $C_2 = 0$, $C_1 \neq 0$ above.

We now list a few elementary algebraic properties of these distributions. They follow straightforwardly from the explicit formula for the CF. More details can be found in [*Samorodnitsky and Taqqu* 1994].

Property A.1 Let $X_1$ and $X_2$ be two independent RVs with $X_i \sim S_{\alpha}(c_i, \beta_i, \kappa_i)$ $(i = 1,2)$ (meaning $X_i$ has the distribution $S_{\alpha}(c_i, \beta_i, \kappa_i)$). Then $X_1 + X_2 \sim S_{\alpha}(c, \beta, \kappa)$ with

$$c^{\alpha} = c_1^{\alpha} + c_2^{\alpha} \ , \ \beta = \frac{\beta_1 c_1^{\alpha} + \beta_2 c_2^{\alpha}}{c_1^{\alpha} + c_2^{\alpha}} \ , \ \kappa = \kappa_1 + \kappa_2 \tag{A.4}$$

Property A.2 Let $X \sim S_{\alpha}(c,\beta,\kappa)$ and let $a$ be any real constant. Then $X + a \sim S_{\alpha}(c,\beta,\kappa + a)$.

Property A.3 Let $X \sim S_{\alpha}(c,\beta,\kappa)$ and let $a$ be any real constant $\neq 0$. Then

$$aX \sim \begin{cases} S_{\alpha}\left(|a|c, \, \mathrm{sgn}(a)\beta, \, a\kappa\right) & ; \ \alpha \neq 1 \\ S_1\left(|a|c, \, \mathrm{sgn}(a)\beta, \, a\kappa - (2/\pi)a \ln|a|\beta\right) & ; \ \alpha = 1 \end{cases} \tag{A.5}$$

In particular, $-X \sim S_{\alpha}(c,-\beta,-\kappa)$.

The moment generating function $a(q) = E[e^{qX}]$ is a quantity of interest to us. We have the following proposition:

Proposition A.1. For a RV $X \sim S_{\alpha}(c,-1,0)$, the function $a(q) = E[e^{qX}]$ ($q$ real and $>0$) is given by

$$a(q) = \begin{cases} \exp\big[-\sec(\pi\alpha/2)c^{\alpha}q^{\alpha}\big] & ; \ \alpha \neq 1 \\ \exp\big[(2/\pi)cq\,\ln q\big] & ; \ \alpha = 1 \end{cases} \tag{A.6}$$

It is ill-defined when $q < 0$.

Existence of the function $a(q)$ is intimately connected with the tail behavior of the distribution as expressed by the next proposition:

Proposition A.2. Let $X \sim S_{\alpha}(c,\beta,\kappa)$ with $0 < \alpha < 2$. Then the decay rate of the tail of the distribution is given by

$$\begin{cases} \lim_{x\to\infty} x^{\alpha}\mathrm{Pr}\{X > x\} = \tfrac{1}{2}A_{\alpha}(1+\beta)c^{\alpha} \\ \lim_{x\to\infty} x^{\alpha}\mathrm{Pr}\{X < -x\} = \tfrac{1}{2}A_{\alpha}(1-\beta)c^{\alpha} \end{cases} \tag{A.7}$$

where

$$A_{\alpha} = \begin{cases} \big[\Gamma(1-\alpha)\cos(\pi\alpha/2)\big]^{-1} & ; \ \alpha \neq 1 \\ 2/\pi & ; \ \alpha = 1 \end{cases} \tag{A.8}$$

In the maximally asymmetric case $\beta = -1$, Proposition A.2 implies that the left tail ($x \to -\infty$) has power law behavior while the right tail ($x \to \infty$) tends to 0 faster than $x^{-\alpha}$. Finding the actual fall-off rate is a somewhat complicated problem [*Zolotarev* 1986] and the final result is quoted in [*Samorodnitsky and Taqqu* 1994]. For the $\alpha = 1$ case the tail behavior is given by Eq. (2.6).

## Appendix B. Moment Function of Log-ID Distributions

This Appendix is devoted to a description of some relevant mathematical properties of the moment function $a(q;L)$ and its analytic continuation, namely the CF $\phi(t;L)$. For simplicity of notation we suppress the $L$-dependence throughout this section.

We consider the functions $a(q)=E[e^{qX}]$ and $\phi(t)=E[e^{itX}]$ for an ID distribution satisfying the conditions (2.2). The RV $X$ is to be identified with the logarithmic rain rate variable $x_L$ introduced in section 2.2. First we derive the integral representation of $\ln a(q)$ given in Eq. (2.16). Starting from Eq. (2.3), namely

$$\ln a(q) = q\kappa + \int_0^{\infty}\left(1 - e^{-qu} - \frac{qu}{1+u^2}\right)dh(u)$$

obtained from analytic continuation of the Lévy canonical representation (2.1), we eliminate the location parameter $\kappa$ using the condition $a(1) = 1$, leading to a simpler form

$$\ln a(q) = \int_0^\infty dh(u)\left[\left(1 - e^{-qu}\right) - q\left(1 - e^{-u}\right)\right]. \tag{B.1}$$

We should emphasize that for the formal analytic continuation to yield a well-defined $a(q)$, it is necessary for the integral to converge. When $q > 0$, the integral indeed converges since $h(u)$ is a non-increasing function of $u$ on $(0,\infty)$. The convergence is not automatic when $q < 0$; to guarantee convergence for all $q$, in addition it is necessary that the function $h(u)$ tend to zero faster than exponential as $u \to \infty$. Otherwise one is led to divergent negative order moments of $e^X$.

The expression (B.1) can be simplified further. An integration by parts (in the Lebesgue-Stieltjes sense) yields

$$\ln a(q) = \left(1 - q\right)h(u)\Big|_{0+}^\infty + qe^{-u}h(u)\Big|_{0+}^\infty - e^{-qu}h(u)\Big|_{0+}^\infty$$
$$+ q\int_0^\infty duh(u)\left[e^{-u} - e^{-qu}\right].$$

The boundary terms at the origin cancel identically even when the function $h(u)$ is singular there. An examination of the first and second boundary terms at infinity show that they vanish individually for all $q$ by virtue of the boundary conditions imposed on $h(u)$ by the Lévy canonical representation. The third term at infinity also vanishes with the additional condition that $h(u)$ tends to zero faster than exponential as $u \to \infty$. Then only the last term survives yielding

$$\ln a(q) = q\int_0^\infty duh(u)\left[e^{-u} - e^{-qu}\right], \tag{B.2}$$

which is Eq. (2.16). Since the spectral function $h(u)$ is non-increasing on $(0,\infty)$ and tends to zero as $u \to \infty$, it follows that $h(u) \geq 0$ on $(0,\infty)$. In the absence of the additional fall-off condition on $h(u)$, the third boundary term at infinity would also survive and would in general diverge when $q < 0$.

It should be noted that $h(u)$ is in general only required to be left-continuous. This allows it to have finite jump discontinuities at a countable set of points (atoms), where the derivative $h'(u)$ has a $\delta$-function singularity. Each atom corresponds to a (suitably shifted and scaled) Poisson component of $X$.

Next, we explore the analytic behavior of the function $a(q)$ defined by Eq. (B.2) which is conveniently rewritten in the form

$$\Lambda(q) \equiv q^{-1}\ln a(q) = \int_0^\infty duh(u)\left[e^{-u} - e^{-qu}\right]. \tag{B.3}$$

Clearly, $\Lambda(1) = 0$ and $\Lambda(0) < 0$. Differentiating under the integral sign we compute the successive derivatives of the function $\Lambda(q)$ with respect to moment order $q$:

$$\Lambda'(q) = \int_0^\infty du \, u h(u) e^{-qu} \ ,$$

$$\Lambda''(q) = -\int_0^\infty du \, u^2 h(u) e^{-qu} \ ,$$

$$...$$

$$\Lambda^{(m)}(q) = (-1)^{m+1} \int_0^\infty du \, u^m h(u) e^{-qu} \ ,$$

(B.4)

and so on. Since the function $h(u)$ satisfies $h(u) \geq 0$ on $(0, \infty)$, it follows that $\Lambda(q)$ is completely monotone, i.e. the successive derivatives of $\Lambda(q)$ alternate in sign: $\Lambda'(q) > 0$, $\Lambda''(q) < 0$, etc. The Taylor series expansion of the function $\Lambda(q)$ at $q=0$, $\Lambda(q) = \sum_{n=1}^\infty (\kappa_n/n!) q^{n-1}$ (if it exists) yields the successive cumulants $\kappa_n$ of the distribution of $X$. Since $a(q) = E[e^{qX}]$, equating like powers of $q$ yields $\kappa_1 = E[X]$, $\kappa_2 = E[X^2] - E^2[X] \equiv \mathrm{Var}[X]$ and so on.

As an example of application of Eq.(B.2) we consider the maximally asymmetric log-Lévy ($\beta = -1$) distribution. This is characterized by the choice $h(u) = C/u^\alpha$, where the stability index $\alpha$ lies in the range $0 < \alpha < 2$. Direct computation yields, for $q > 0$,

$$\ln a(q) = Cq \int_0^\infty du \, u^{-\alpha} \left[ e^{-u} - e^{-qu} \right]$$

$$= C \left( q - q^\alpha \right) \int_0^\infty du \, u^{-\alpha} e^{-u}$$

$$= C\Gamma(1-\alpha)\left( q - q^\alpha \right)$$

The special case $\alpha = 1$ can be easily accommodated by a limiting procedure:

$$\ln a(q) = C \, \mathrm{Lim}_{\alpha \to 1} \, \Gamma(1-\alpha)\left( q - q^\alpha \right)$$

$$= C \, \mathrm{Lim}_{\alpha \to 1} \, \Gamma(2-\alpha) \frac{\left( q - q^\alpha \right)}{1-\alpha}$$

$$= C \, \mathrm{Lim}_{\alpha \to 1} \frac{\left( q - q^\alpha \right)}{1-\alpha}$$

$$= Cq \ln q$$

in agreement with Proposition A.1 above (up to a nontrivial centering term linear in $q$ when $\alpha \neq 1$). The computation fails when $q < 0$ since the original integral diverges. Nonexistence of the negative order moments of $e^X$ is consistent with the distribution of $X$ having a slowly decaying power law tail along the negative axis.

Explicit computation of the function $\Lambda(q)$ at various $q$ from the moments of the precipitation data suggest that one might try to represent it as a Taylor series at $q = 0$ in the form (taking into account the fact that $\Lambda(q)$ has a simple zero at $q = 1$) :

$$\Lambda(q) = (q-1)\left[c_0 + c_1 q + c_2 q^2 + c_3 q^3 + ...\right] \qquad (B.5)$$

These coefficients are simply certain linear combinations of the various cumulants $\kappa_n$ introduced above: $c_0 = -\kappa_1$, $c_0 - c_1 = \frac{1}{2}\kappa_2$, etc. They can be explicitly calculated for a specified Lévy spectral function $h(u)$. The inequalities $(-1)^m \kappa_m > 0$ $(m = 1, 2, ...)$ then imply that the coefficients $c_0, c_1, c_2 ...$ must satisfy the sequence of inequalities $c_0 > 0$, $c_1 < c_0$, $c_2 > c_1$ and so on.

For the new distribution defined by Eq. (4.3) the integral (B.3) converges for all $q$, both positive and negative. We have, when $q > 0$,

$$\begin{aligned}
\Lambda(q) &= (2c/\pi)\int_0^b du\, u^{-1}\left[e^{-u} - e^{-qu}\right] \\
&= (2c/\pi)\left(\int_0^\infty - \int_b^\infty\right) du\, u^{-1}\left[e^{-u} - e^{-qu}\right] \\
&= (2c/\pi)\left[\ln q + E_1(bq) - E_1(b)\right] \\
&= (2c/\pi)\left[E_1(z) + \ln z\right]_b^{bq}
\end{aligned} \qquad (B.6)$$

where $E_1(z) = \int_z^\infty dt(e^{-t}/t)$ is one of the exponential integral functions of a complex variable $z$ [*Abramowitz and Stegun* 1972]. $E_1(z)$ has a logarithmic branch point at the origin and is defined as an analytic function in the complex $z$–plane ($|\arg z| < \pi$) cut along the negative real axis. It can be represented by a series expansion

$$E_1(z) = -\gamma - \ln z - \sum_{n=1}^\infty \frac{(-z)^n}{n \cdot n!} \qquad \left(|\arg z| \neq \pi\right)$$

where $\gamma$ is Euler's constant. The combination $\mathrm{Ein}(z) \equiv E_1(z) + \gamma + \ln z$ which appears in the final step of Eq.(B.6), is thus analytic in the entire complex $z$-plane since the branch point singularities of the individual terms cancel each other out. This allows one to analytically extend $\Lambda(q)$ to $q < 0$ as follows (even though the above formal derivation fails). One defines the function $\mathrm{Ei}(x)$ of a real variable $x$ as [*Abramowitz and Stegun* 1972]

$$\mathrm{Ei}(x) = \begin{cases} -\int_{-x}^\infty dt\left(e^{-t}/t\right) = -E_1(-x) & (x < 0) \\[2mm] -\,\mathrm{P}\int_{-x}^\infty dt\left(e^{-t}/t\right) & (x > 0) \end{cases}$$

Here P denotes the integral defined by the Cauchy principal value prescription at the origin where the integrand encounters a singularity. [This is in complete analogy with the elementary results $\ln x = -\int_x^1 dt/t$ $(x > 0)$ and $\ln |x| = -\mathrm{P}\int_x^1 dt/t$ $(x < 0)$]. Like the logarithmic function, $E_1(z)$ has a finite discontinuity $2\pi i$ across the branch cut along the negative real axis with the assigned values

$$E_1(-x \pm i0) = -\mathrm{Ei}(x) \mp \pi i$$

so that

$$-\mathrm{Ei}(x) = \tfrac{1}{2}\big[E_1(-x+i0) + E_1(-x-i0)\big] \qquad (x > 0)$$

The function $\mathrm{Ei}(x)$ can then be expressed in the form

$$\mathrm{Ei}(x) = \begin{cases} \gamma + \ln|x| - \mathrm{Ein}(-x) & (x < 0) \\ \gamma + \ln x - \mathrm{Ein}(-x) & (x > 0) \end{cases}$$

Returning to the evaluation of $\Lambda(q)$, we can therefore combine both cases $q>0$ and $q<0$ into a single expression:

$$\Lambda(q) = (2c/\pi)\big[\ln|q| - \mathrm{Ei}(-bq) + \mathrm{Ei}(-b)\big] \qquad (B.7)$$

which yields Eq.(4.4). Its derivatives are expressible in terms of elementary functions: $\Lambda'(q) = (2c/\pi)(1-e^{-bq})/q$ and so on. Truncation of the Lévy spectral integral has the effect of rendering all the moments of $e^X$ finite and well defined.

The CF of the new distribution is given by the integral representation

$$\ln \phi(t) = it \int_0^\infty du\, h(u)\big[e^{-u} - e^{-itu}\big]$$
$$= (2i/\pi)ct \int_0^b du\, u^{-1}\big[e^{-u} - e^{-itu}\big].$$

Explicit evaluation yields Eq. (4.6):

$$\ln \phi(t) = (2c/\pi)\Big\{-|t|\mathrm{Si}(b|t|) + it\big[\ln|t| - \mathrm{Ci}(b|t|) - E_1(b)\big]\Big\},$$

where $\mathrm{Si}(z) = \int_0^z dt\,(\sin t/t)$ and $\mathrm{Ci}(z) = -\int_z^\infty dt\,(\cos t/t)$ denote the sine and cosine integral functions respectively [*Abramowitz and Stegun* 1972] and are related to the exponential integral function $E_1(z)$ through the formula

$$E_1(iz) = -\mathrm{Ci}(z) + i\,\mathrm{Si}(z) - i\pi/2$$

While $\mathrm{Si}(z)$ is an analytic function, $\mathrm{Ci}(z)$ inherits a logarithmic branch point singularity at the origin from $E_1(z)$ and can be expressed in the form

$$\mathrm{Ci}(z) = \gamma + \ln z - \mathrm{Cin}(z)$$

where $\mathrm{Cin}(z) = \int_0^z dt\,(1 - \cos t)/t$ is an entire function of $z$ in the complex plane.

## Appendix C. Computation of the Function $g(x_L; c(L), b(L))$

In this Appendix we outline some of the details involved in the computation of the pdf of the logarithmic rain rate variable $x_L$. As before we suppress the $L$-dependence throughout this Appendix.

The pdf $g(x;c,b)$ is computed by numerically evaluating the Fourier transform integral

$$g(x;c,b) = \frac{1}{2\pi} \int_{-\infty}^{\infty} dt\, e^{-itx} \phi(t)$$

In view of the reality condition $\phi(-t) = \bar{\phi}(t)$, it can be expressed in the form

$$g(x;c,b) = \frac{1}{\pi} \int_0^{\infty} dt\, \exp\left[-(2c/\pi)t\,\mathrm{Si}(bt)\right]\cos\left[xt - \Theta(t;c,b)\right] \tag{C.1}$$

where

$$\Theta(t;c,b) = (2c/\pi)t\left[\ln t - \mathrm{Ci}(bt) - E_1(b)\right] \tag{C.2}$$

Note that the functions $\mathrm{Ci}(z)$ and $\mathrm{Si}(z)$ appearing in the integrand in (C.1) need to be evaluated only for positive arguments. The necessary numerical routine is provided by *Press et al.* [1995].

The Fourier cosine integral (C.1) is computed for a specified value of $c$ and $b$ by utilizing a numerical implementation based on Fast Fourier Transform (FFT) algorithm as described by *Press et al.* [1995]. The routine was tested with several known examples – the normal, the Cauchy and the $S_1(c,-1,0)$ Lévy stable distributions. In the first two cases simple analytic results are available. In the last case the computed pdf was checked against that obtained from an alternative integral representation of the pdf of stable distributions due to *Nolan* [1997], which has the numerical advantage of having a non-oscillatory integrand. For the $S_1(1,\beta,0)$ distribution Nolan's representation of the pdf reads (in the case $\beta \neq 0$)

$$v(x;1,\beta) = \frac{1}{2|\beta|} e^{-\pi x/2\beta} \int_{-\pi/2}^{\pi/2} V(\theta;1,\beta) \exp\left[-e^{-\pi x/2\beta} V(\theta;1,\beta)\right] d\theta$$

where

$$V(\theta;1,\beta) = \frac{2}{\pi} \frac{\left(\tfrac{\pi}{2} + \beta\theta\right)}{\cos\theta} \exp\left[\tfrac{1}{\beta}\left(\tfrac{\pi}{2} + \beta\theta\right)\tan\theta\right]$$

No explicit analytical results are available for checking the computation of the new pdf $g(x;c,b)$. However, a powerful consistency check is provided by the fact that $g(x;c,b)$ satisfies the scaling identity

$$g(\lambda x; \lambda c, \lambda b) = \lambda^{-1} g(x + \xi; c, b)$$

with a shift $\xi$ given by

$$\xi = c\left[ E_1(\lambda b) - E_1(b) - \ln \lambda \right]$$

The equality was verified by explicit computation from our numerical algorithm.

## References.

**Figure Captions**

Figure 1.   Plots of $\Lambda(q) = q^{-1}\ln a(q;L)$ vs. $q$ from data (open circles) compared with predictions from the proposed distribution, Eq.(4.4) and the parameters listed in Table I (solid line) for (a) $L = 2$ km and (b) $L = 128$ km for MIT Cruise 3 and TOGA Cruise 3. The level of agreement is similar for all other $L$.

Figure 2.   Plots of the pdf $g(x_L;c(L), b(L))$ vs. $x_L$ superimposed on the observed rain rate histograms for MIT Cruise 3 and TOGA Cruise 3 data at different spatial scales: (a) $L = 2$ km, (b) $L = 8$ km, (c) $L = 32$ km and (d) $L = 128$ km. The pdf (solid curve) is scaled so that the area under the curve equals the area of the histogram between the observed maximum and minimum rain rates. Agreement at the other explored scales $L = 4$ km, 16 km and 64 km (not shown) are also deemed satisfactory.

Figure 3. Quantile plots of data vs. model for MIT Cruise 3 and TOGA Cruise 3 data at different spatial scales: (a) $L = 2$ km, (b) $L = 8$ km, (c) $L = 32$ km and (d) $L = 128$ km. (The straight line represents the diagonal $y = x$).

Figure 4. Linear regression of $c(L)$ (solid circles), $\ln p(L)$ (solid triangles) and $b(L)$ (open circles) against $\ln L$ between $L = 2$ km and $L = 128$ km for MIT Cruise 3 and TOGA Cruise 3.

Figure 5.  Log-log plots of $m(q;L)$ vs. $L$ for selected values of $q$ illustrating power law scaling of the moments for MIT Cruise 3 and TOGA Cruise 3.

Figure 6.   The observed scaling exponents $\eta(q)$ (solid circles) compared with the predictions from the new (solid line) and the lognormal (dashed line) models for (a) MIT Cruise 3 and (b) TOGA Cruise 3.

<u>Table 1</u>

## Model Parameters $p(L)$, $c(L)$, $b(L)$ and related quantities for Cruise 3*

### (a) MIT Radar

| $L$ (km) | 2 | 4 | 8 | 16 | 32 | 64 | 128 |
|---|---|---|---|---|---|---|---|
| $p(L)$ | 0.106 | 0.139 | 0.198 | 0.303 | 0.472 | 0.683 | 0.858 |
| $c(L)$ | 3.0 | 2.8 | 2.6 | 2.4 | 2.2 | 2.0 | 1.7 |
| $b(L)$ | 1.0 | 1.6 | 2.2 | 2.8 | 3.3 | 3.9 | 4.6 |
| $c_0(L)$(model) | 1.52 | 2.02 | 2.32 | 2.48 | 2.49 | 2.47 | 2.28 |
| $c_1(L)$(model) | −0.39 | −0.83 | −1.32 | −1.80 | −2.13 | −2.49 | −2.70 |
| $c_0(L)$(fit) | 1.664(0.002) | 1.928(0.001) | 2.171(0.002) | 2.363(0.005) | 2.480(0.008) | 2.458(0.011) | 2.228(0.012) |
| $c_1(L)$(fit) | −0.073(0.013) | −0.398(0.005) | −0.777(0.009) | −1.129(0.025) | −1.405(0.043) | −1.551(0.059) | −1.496(0.067) |

### (b) TOGA Radar

| $L$ (km) | 2 | 4 | 8 | 16 | 32 | 64 | 128 |
|---|---|---|---|---|---|---|---|
| $p(L)$ | 0.084 | 0.109 | 0.155 | 0.239 | 0.374 | 0.546 | 0.730 |
| $c(L)$ | 3.5 | 3.2 | 2.8 | 2.5 | 2.2 | 1.9 | 1.7 |
| $b(L)$ | 0.9 | 1.3 | 1.9 | 2.8 | 3.6 | 4.4 | 5.1 |
| $c_0(L)$(model) | 1.63 | 1.99 | 2.27 | 2.58 | 2.61 | 2.49 | 2.39 |
| $c_1(L)$(model) | −0.37 | −0.66 | −1.11 | −1.87 | −2.43 | −2.83 | −3.13 |
| $c_0(L)$(fit) | 1.647(0.002) | 1.906(0.001) | 2.151(0.002) | 2.346(0.005) | 2.444(0.007) | 2.418(0.011) | 2.413(0.014) |
| $c_1(L)$(fit) | −0.044(0.011) | −0.375(0.004) | −0.763(0.010) | −1.124(0.026) | −1.358(0.040) | −1.552(0.061) | −1.708(0.076) |

* The quantities within parentheses are the standard errors.

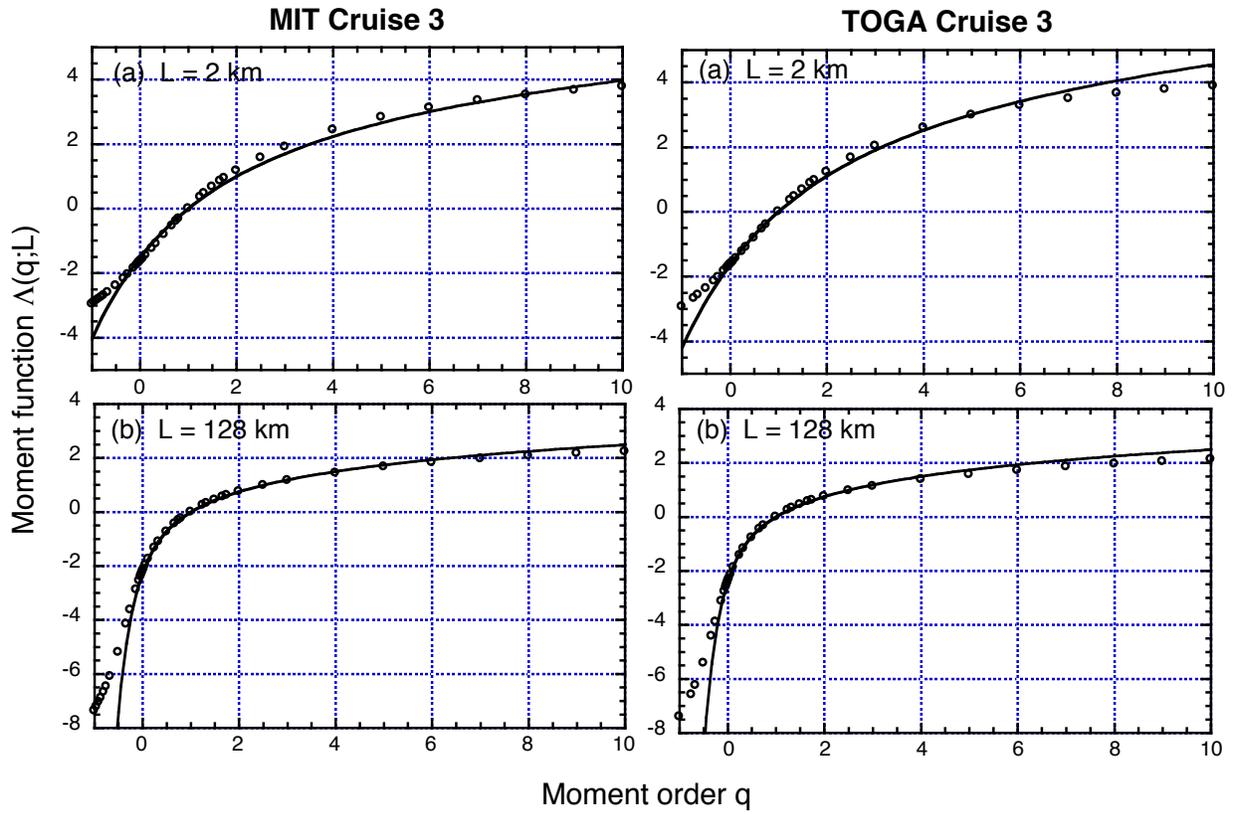

**Figure 1**

**MIT**                    **TOGA**

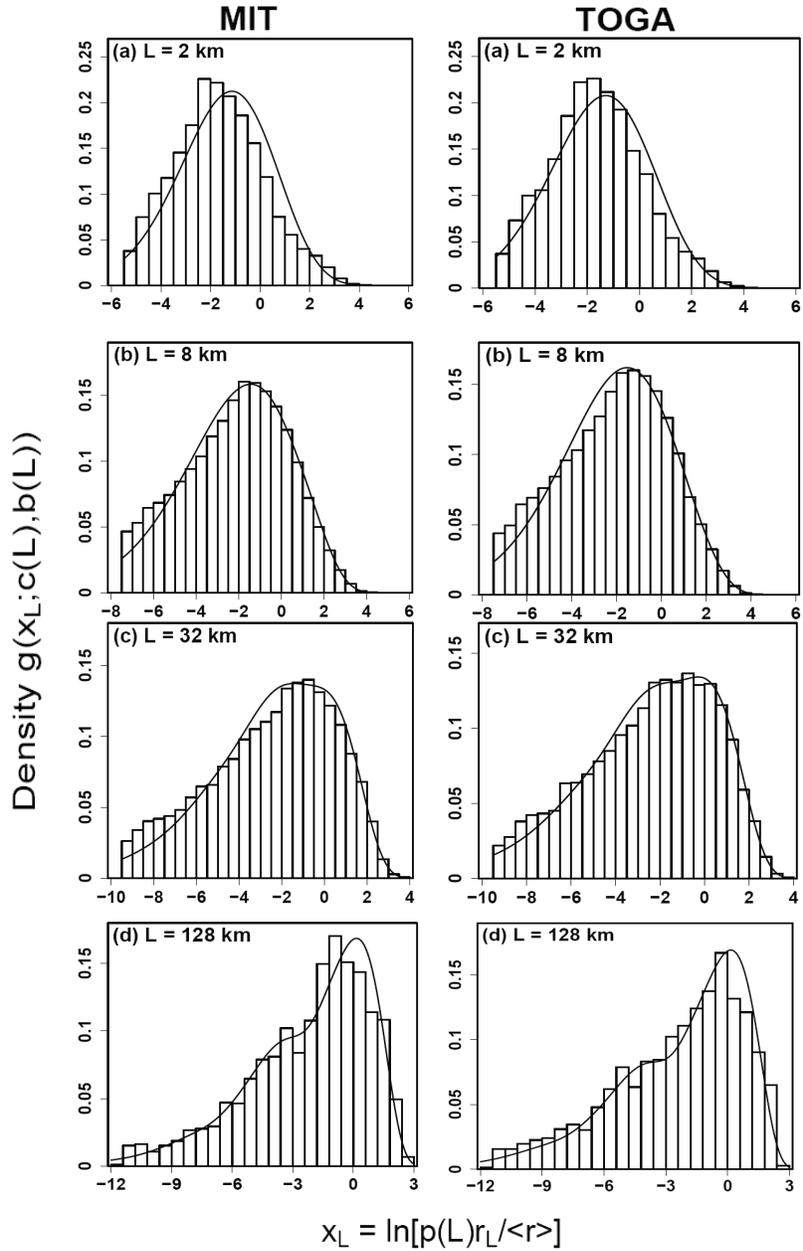

$$x_L = \ln[p(L)r_L/\langle r \rangle]$$

**Figure 2**

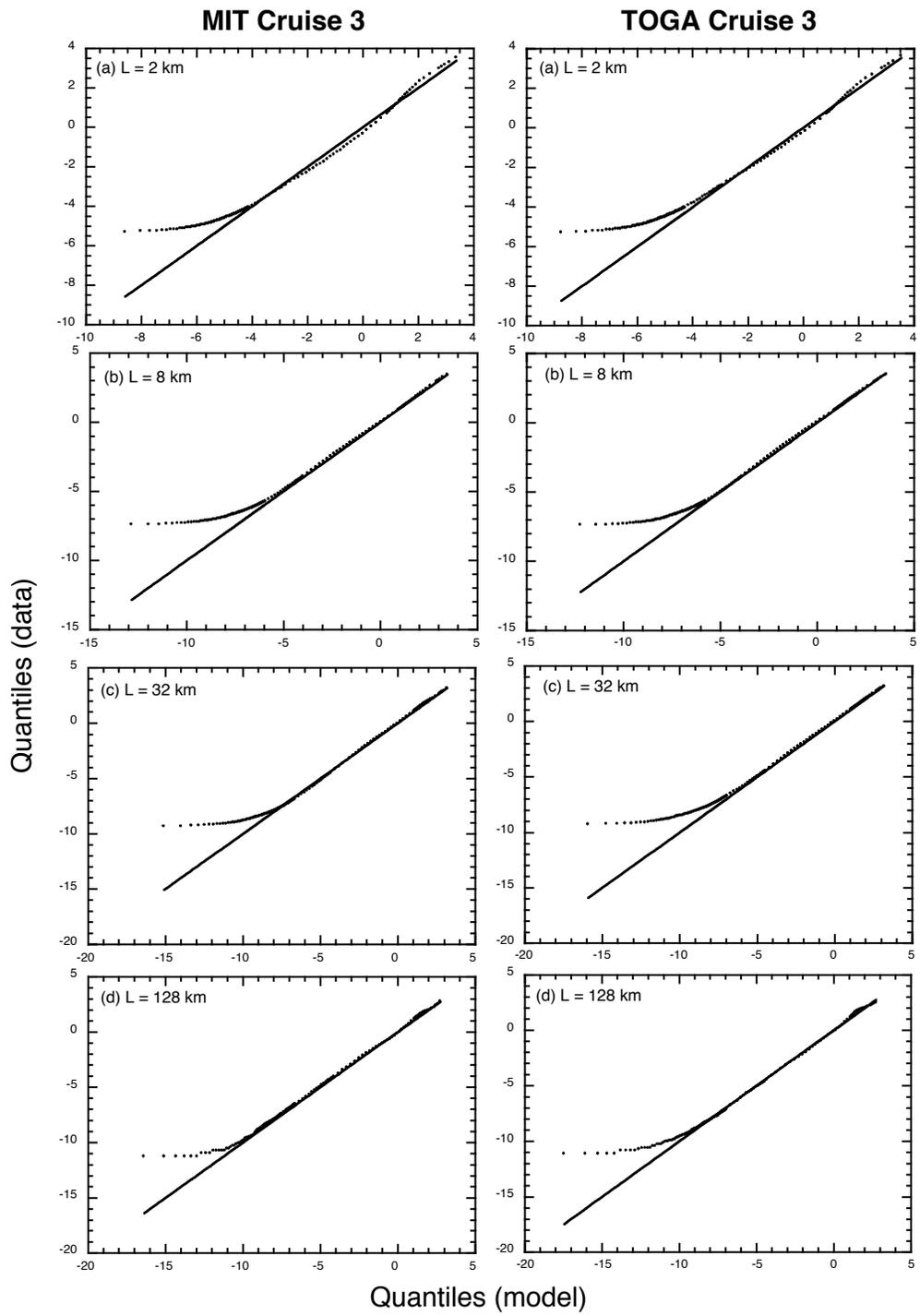

**Figure 3**

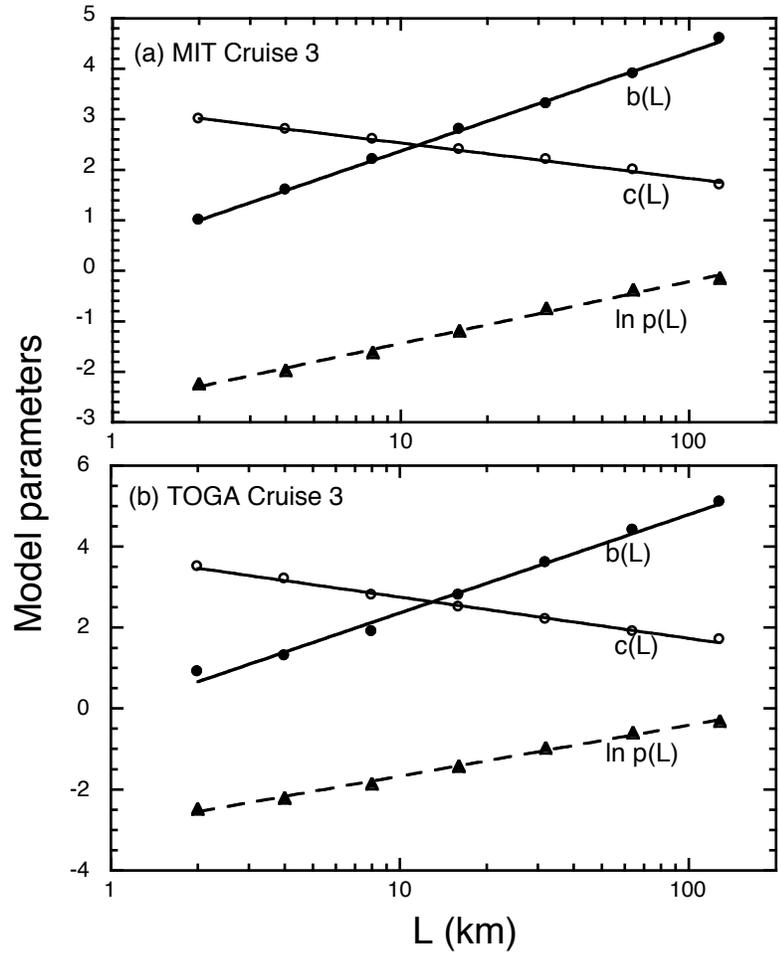

**Figure 4**

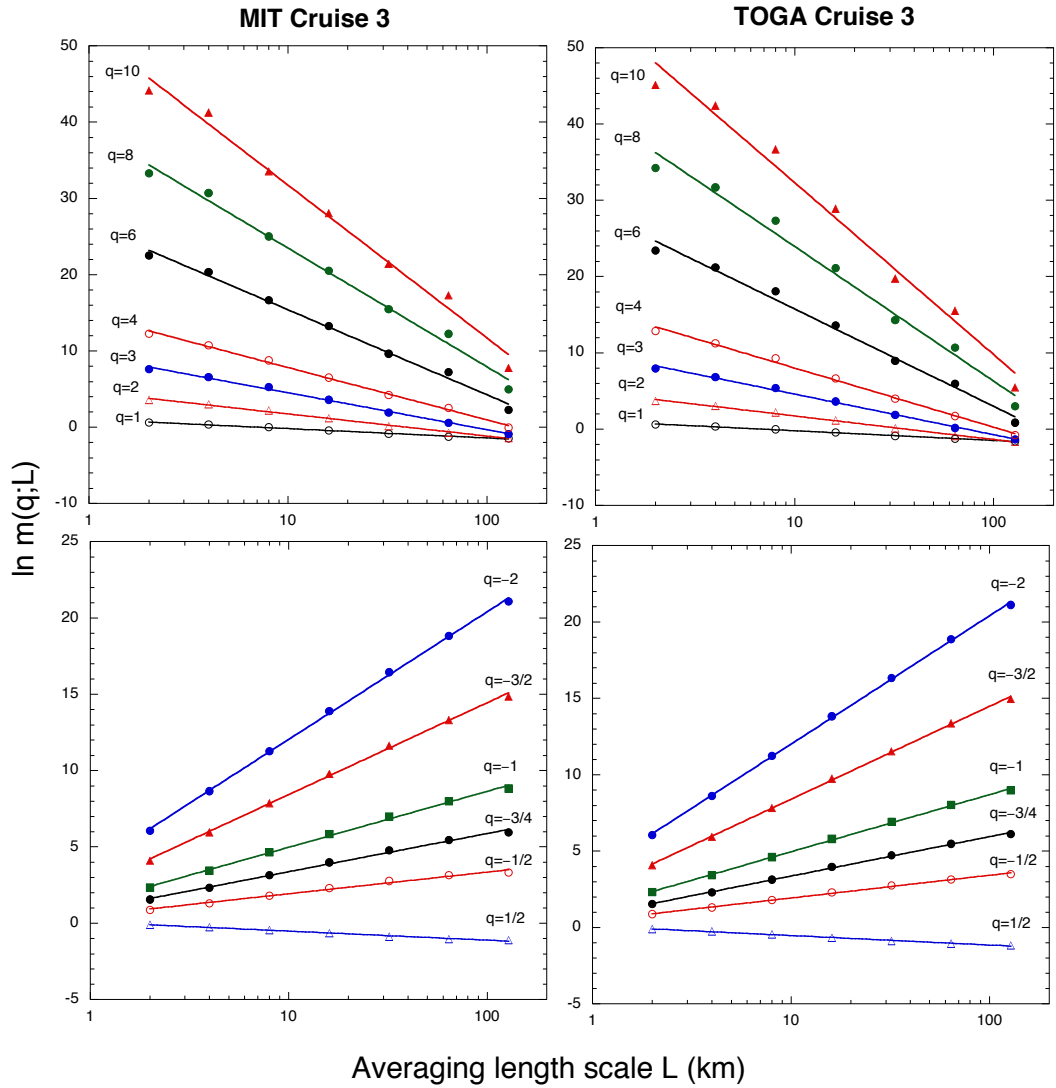

**Figure 5**

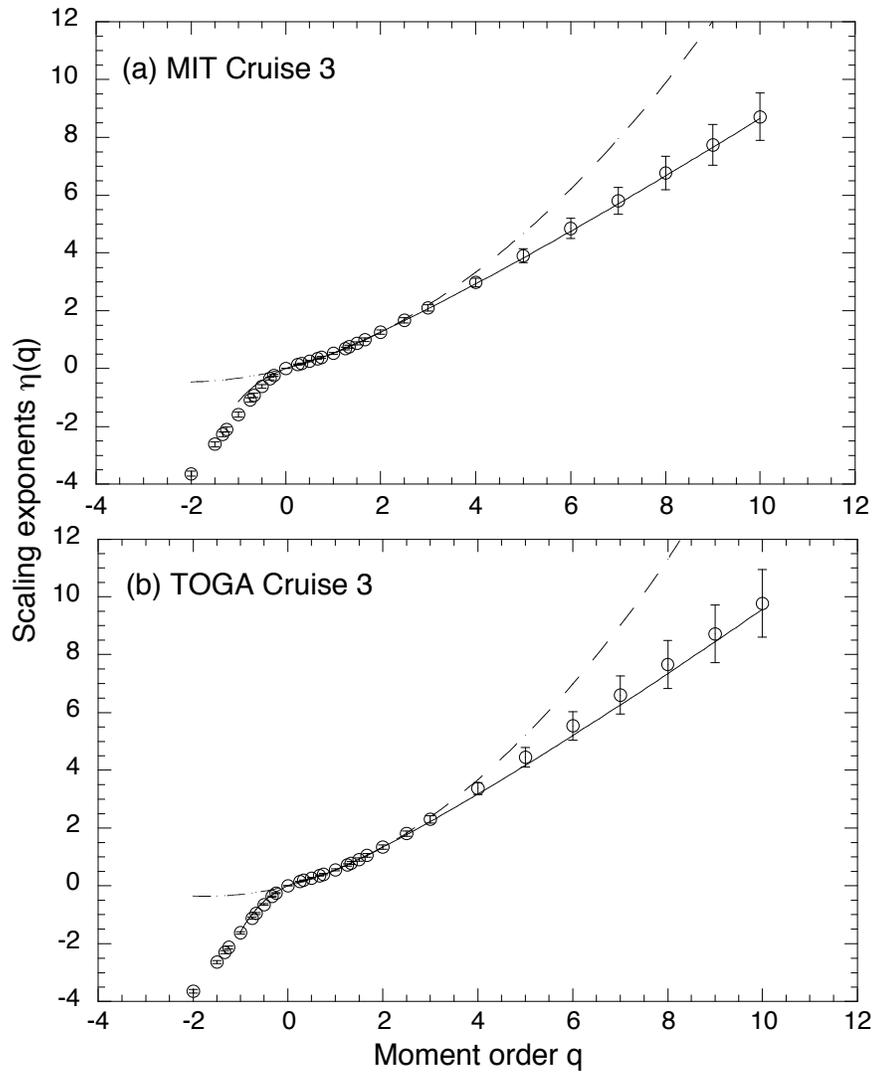

**Figure 6**